# Multiple Andreev Reflections Spectroscopy of Superconducting LiFeAs Single Crystals: Anisotropy and Temperature Behavior of the Order Parameters

S. A. Kuzmichev[1], T. E. Kuzmicheva[2], A. I. Boltalin[1], and I. V. Morozov[1]

[1] *Moscow State University, Moscow, 119991 Russia*, e-mail: *kuzmichev at mig.phys.msu.ru*

[2] *Lebedev Physical Institute, Russian Academy of Sciences, Moscow, 119991 Russia*

The superconducting state of LiFeAs single crystals with the maximum critical temperature $T_c \approx 17$ K in the 111 family has been studied in detail by multiple Andreev reflections (MAR) spectroscopy implemented by the break-junction technique. The three superconducting gaps, $\Delta_\Gamma = 5.1 - 6.5$ meV, $\Delta_L = 3.8-4.8$ meV, and $\Delta_S = 0.9 - 1.9$ meV (at $T \ll T_c$), as well as their temperature dependences, have been directly determined in a tunneling experiment with these samples. The anisotropy degrees of the order parameters in the $k$ space have been estimated as <8, ~12, and ~20%, respectively. Andreev spectra have been fitted within the extended Kummel–Gunsenheimer–Nikolsky model with allowance for anisotropy. The relative electron-boson coupling constants in LiFeAs have been determined by approximating the $\Delta(T)$ dependences by the system of the two-band Moskalenko and Suhl equations. It has been shown that the densities of states in bands forming $\Delta_\Gamma$ and $\Delta_L$ are approximately the same, intraband pairing dominates in this case, and the interband coupling constants are related as $\lambda_{\Gamma L} \approx \lambda_{L\Gamma} \ll \lambda_{S\Gamma}, \lambda_{SL}$.

Iron-containing superconductors [1] have been studied for almost five years. However, interest in these compounds remains. The main unanswered question is: What is the mechanism of the formation of Cooper pairs responsible for such high critical temperatures in superconductors with magnetic atoms in the structure? Although the iron isotopic effect was observed experimentally [2] with the coefficient $\alpha \approx 0.4 < 0.5$, the strong electron-phonon interaction [3] does not describe the observed $T_c$ values [4]. In view of the closeness of the main antiferromagnetic state with the spin density wave and wave vector $Q_{AFM} = (\pi/a, \pi/a)$ [5] and the congruence of the Fermi surfaces in the Γ–M direction (the so-called nesting on the $Q_{\Gamma-M} \equiv Q_{AFM}$ vector) [6], the dynamic magnetic susceptibility has a peak (so-called magnetic resonance) at the energy $E_{res} \approx 1.5\Delta_L$ according to calculations reported in [7], and the ratio $E_{res}/k_B T_c \approx 5.5$ according to the spectroscopic data on the inelastic scattering on neutrons presented in [8]. The facts listed above initiate the theoretical idea that spin fluctuations play a significant role in the formation of Cooper pairs, which are responsible for the strong interband interaction and for change in the sign of the order parameter between the sheets of the Fermi surface ($s^{\pm}$-model) [9].

Studies of the superconducting properties of LiFeAs are of key importance in the search for the answer to the formulated question. The layered structure of this material, as well as other iron-containing superconductors, is a stack of FeAs blocks separated by spacers (double Li layers in this case [10]) along the $c$ direction. The band structure, density of states, and structure of the Fermi surfaces of LiFeAs are also similar to the characteristics of other superconducting iron arsenides [11–14]. However, in contrast to other pnictides, LiFeAs undergoes neither structural nor magnetic transitions [15] and superconducts in the stoichiometric composition [10, 16, 17]. The most surprising fact is the absence of nesting in LiFeAs, while the appearance of superconductivity is due to the presence of the van Hove singularity near the Fermi level of the undoped compound [16]. Does this imply that the mechanism of superconductivity in LiFeAs is different from that in other pnictides? On one hand, an "extended" $s^{\pm}$ model was proposed in several theoretical works [13, 18–20] to explain superconductivity in LiFeAs. On the other hand, it was shown that a system with interaction through spin fluctuations is unstable with respect to scattering on impurities. The model of an isotropic two-gap superconductor with the order parameters of the same sign, where pairs are formed owing to interaction through orbital fluctuations ($s^{++}$), was suggested [21]. In addition, the impossibility of the implementation of the $s^{\pm}$-system in LiFeAs was proved in a number of theoretical calculations [14, 22]. This problem can be clarified by detailed experimental studies of the superconducting state of LiFeAs: the determination of the number and values of superconducting gaps $\Delta$, their distribution in the directions in the k space, the relation of $2\Delta$ to the energy of magnetic resonance, the temperature dependences $\Delta(T)$, and the corresponding electron-boson coupling constants. Unfortunately, in view of difficulties in handling of LiFeAs samples (the material is strongly hygroscopic because lithium atoms which are quite weakly bound in the layers, have significant chemical activity), the existing data [16, 23–31] (see also Table 1 in [32]) are very contradictory.

The existence of two anisotropic superconducting gaps – $\Delta_L$ = 2–3.5 meV in electron bands (M point) and $\Delta_S$ = 1.5–2.5 meV in hole bands (Γ point) – was demonstrated by Borisenko et al. [16] in one of the first works on angle-resolved photoemission spectroscopy (ARPES) on samples grown by the same method as those used in this work. More recently, the same team presented refined data [23] and reported the observation of a large isotropic gap $\Delta_h^{(in)} \approx 6$ meV on the inner hole cylinder of the Fermi surface (below, we denote this gap as $\Delta_\Gamma$), anisotropic small gap $\Delta_h^{(ex)} \approx [3.4 + 0.5 \cos(4\theta)]$ meV on the outer hole cylinder, and "electron" gap $\Delta_e = [3.6 + 0.5 \cos(4\theta)]$ meV. Furthermore, according to the data reported in [23], the "hole" gap $\Delta_h^{(ex)}$ has the minimum in the Γ–M direction (the $s^{\pm}$-model implies an opposite relation). Thus, although ARPES is insensitive to the phase of the order parameter, the resulting angular distribution of the gap is in favor of the $s^{++}$ mechanism [23]. The large gap $\Delta_\Gamma$ = 5–6 meV was observed in ARPES studies [24, 29] and in scanning tunneling microscopy (STM) studies [26, 31]. However, there are significant discrepancies between experimental data on the angular distribution of this gap in the k space: the gap is isotropic according to the data reported in [24], whereas its significant anisotropy was stated in [25, 26]. An even larger spread is observed in the experimental data for small gaps. In addition to the large gap $\Delta_\Gamma \approx 6$ meV, the authors of [29] observed a second gap with an amplitude of 3.4 meV without anisotropy. At the same time, the STM results [26] indicate the strong anisotropy (~40%) of the small gap $\Delta_S \approx 2.5$ meV. The observation of the four-gap superconductivity in LiFeAs was stated in [24], where it was shown that condensates with the large gap $\Delta_\Gamma \approx 5.0$ meV (on the inner cylinder) and the small gap of 2.2–3 meV (on the outer cylinder, anisotropy of ~30%) are formed at the $\Gamma$ point and condensates with the gaps $\Delta_L$ = 3–4.2 meV (on the outer cylinder, anisotropy of ~30%) and ~2.9 meV (on the inner cylinder, weak anisotropy of ~5%) are formed at the M point. A strongly anisotropic (~40%) small gap with an average value of 2.5 meV was also observed in [25]. The optical measurements [28] demonstrated the existence of the gap $\Delta_L \approx 3.3$ meV and the smallest gap $\Delta_S \approx 1.6$ meV, whose existence was also confirmed in the point-contact spectroscopy study [27]. The common conclusion of the experimental studies of the energy parameters of LiFeAs is that the characteristic ratio of the Bardeen–Cooper–Schrieffer (BCS) theory, $2\Delta_\Gamma/k_BT_c$, lies in the range of 6–8.5 and $2\Delta_L/k_BT_c \approx 4.8$–5.5 for the condensate in electron bands.

In this work, we present the dynamic conductance spectra of symmetric superconductor–normal metal–superconductor (SNS) contacts obtained by the break-junction technique in LiFeAs single crystals with $T_c$ = 16–17 K. Our experimental results indicate the presence of three-gap superconductivity. The spectra of SNS Andreev junctions were obtained for the first time, which exhibit pronounced, but split features corresponding to the large gaps $\Delta_\Gamma$ = 5.1–6.5 meV with a weak anisotropy (<8%) and $\Delta_L$ = 3.8–4.8 meV with a fairly strong anisotropy (~12%), as well as a small gap $\Delta_S$ = 0.9–1.9 meV for which the splitting of features and, correspondingly, anisotropy are significant and are no less than 20%. It was shown that the temperature dependences obtained for the gaps can be described by the system of the two-band Moskalenko and Suhl equations [33] (with the renormalized BCS integral). Since the temperature dependences of large gaps $\Delta_{\Gamma,L}(T)$ are almost identical and, therefore, the densities of states in the bands are approximately the same, $N_L \approx N_\Gamma$, these bands can be considered as an effective unified band with $\Delta_L^{eff} \approx (\Delta_\Gamma + \Delta_L)/2$. Thus, the two-gap approach is applicable for the description of the temperature dependence of the order parameters $\Delta_i$. The relative intra- and interband electron-boson coupling constants are determined from our experimental data. It is shown that the interband interaction in LiFeAs is much weaker than the intraband interaction. The comparison between the gap values obtained and the magnetic resonance energy is also presented.

We used LiFeAs single crystals with bulk critical temperatures $T_c^{bulk}$ = 16–17 K and whose synthesis and characterization were described in detail in [34]. The superconducting properties were studied by Andreev spectroscopy [35]. This method is based on the effect of multiple Andreev reflections in Sharvin-type SNS nanojunctions [36] (their diameter $a$ is smaller than the quasiparticle mean free path $l$), so-called ballistic junctions. A quasiparticle accelerated owing to the applied bias voltage passes through the $N$ layer, which behaves as a metal. Reflecting from the metal–superconductor interfaces, it changes the sign of the charge and both components of the velocity, acquiring the energy eV in each pass until its energy becomes sufficient to enter the quasiparticle branch of the conduction band of the superconductor, $E_{min} \geq E_F + \Delta$. Thus, at biases $V_n = 2\Delta/en$ (where n = 1, 2, …), the dynamic conductance spectrum of such a junction will exhibit minima constituting a subharmonic gap structure (SGS) [37–39]. It is obvious that $V_n \rightarrow 0$ at $n \rightarrow \infty$ and a significant excess current appears at low bias voltages owing to increasing efficiency of Andreev transport. This current is manifested in the current-voltage characteristic as a typical linear segment with a high steepness called foot. The presence of a foot on the current-voltage characteristic is the most important attribute of the SNS junction.

According to the theory [39] describing high-transparency SNS junctions, the number of observed Andreev peculiarities corresponds to the ratio of the mean free path to the diameter of the junction, $l/a$. The authors of [39] showed that the positions of the minima of the subharmonic gap structure correspond to the order parameter up to $T_c$. Thus, the gap width can be determined as $2\Delta = eV_n \cdot n$ at any temperatures in the superconducting state exists. The last circumstance is very important because it allows the determination of $\Delta$ directly from experimental spectra without fitting with many varying parameters (seven and eleven for the two- and three-gap cases, respectively). Nevertheless, for the features of the subharmonic gap structure caused by the anisotropic order parameter, it is reasonable to compare the experimental and calculated dI/dV spectra of the SNS Andreev junction because this makes it possible to determine both the cause and magnitude of anisotropy if the anisotropy of $\Delta$ is ≤ 45%. This fitting is particularly simple if the Andreev current flows primarily in the $c$ direction because the contributions to the conductivity of the SNS junction from the varying (in the k or real space) amplitude of the order parameter $\Delta$ can be considered as completely independent (parallel), which was done in our calculations. Several subharmonic gap structures (for each of the gaps) will be observed in the spectrum of a multiband superconductor.

Symmetric SNS junctions were created in the samples by the break-junction technique [40]. A rectangular sample $3 \times 1.5 \times 0.2$ mm$^3$ in size was fixed on a spring measuring table by a liquid indium-gallium solder by the four contact connection. In order to prevent the degradation of LiFeAs in open air, the sample was mounted in an argon atmosphere. At the mechanical deflection of the table preliminarily cooled to T = 4.2 K, a microcrack appeared in the sample. As was shown in [16], the LiFeAs layered single crystal is cleaved just along the ab direction between lithium planes. Correspondingly, the microcrack creates an area of a normal metal separating two superconducting banks, i.e., forms the SNS junction in the *c* direction. Massive superconducting banks, as well as the remoteness of current contacts of the sample, ensure good heat removal from the measurement region (junction). The location of the microcrack deep in the sample prevents the penetration of impurities on the surfaces of the cryogenic cleavage, leaving them as clean as possible.

In terms of the resistivity of our samples in the normal state $\rho_N(20\ \text{K}) = (1-5)\times 10^5$ mΩ·cm, mean free path $l = 4\text{–}5$ nm [41], and typical resistance of our SNS junctions $R_N \sim 10$ Ω, the diameter of the junction can be calculated by the Sharvin formula [36]

$$a = \sqrt{\frac{4}{3\pi}\frac{\rho_N l}{R}} \approx 1-3\ \text{nm} < l.$$

Thus, although the difference between *a* and *l* is small, the resulting break junctions are in the ballistic regime. One or two Andreev features are expected in the spectra.

In view of the pronounced stratification of the crystal lattice of LiFeAs, step-and-terrace structures are formed on the surface of cryogenic cleavages (the height of the steps is a multiple of the unit cell parameter *c*). They can serve as stack junctions, i.e., natural structures of the S–N–S–N–…–S type [42]. Such a stack is electrically equivalent to several in-series connected identical SNS junctions. Consequently, bias voltages of any features (reflecting the bulk properties of the material) in the current-voltage characteristic and dI(V)/dV spectrum will be scaled by a factor of N, where N is the number of junctions in the stack [43]. This means the intrinsic multiple Andreev reflections effect, which was observed for the first time on Bi-2201 [44] and, then, on all layered materials studied. Normalizing the stack dI/dV characteristics to the spectrum of a single SNS junction, we can determine the number N and gap widths. Studies of stack junctions guarantee the measurement of bulk superconducting gaps. It was shown in [43] that the contribution of inhomogeneities introduced by the surface of the cryogenic cleavage decreases with an increase in N and the gap peculiarities on the spectra of the dynamic conductance become sharper, indicating an increase in the accuracy of the determination of the gaps by a factor of about N. The results presented below were obtained by two methods:

Andreev spectroscopy of single SNS junctions and intrinsic Andreev spectroscopy of S–N–S–N–…–S stacks.

Figure 1 shows the current-voltage characteristics in comparison with the dynamic conductance spectra for the SNS Andreev junctions. The lines are vertically shifted with respect to each other for convenience. The three upper spectra connected by a vertical arrow (on the right) correspond to single SNS Andreev junctions (at T = 4.2 K) obtained by successive mechanical readjustment on the LFA12 sample: LFA12_d8, LFA12_b, and LFA12_f2. The current-voltage characteristics of these junctions (represented by the same line types as the respective derivatives) are strongly nonlinear. In particular, the pronounced segment of the excess current at small biases (foot) indicates that junctions are in the Andreev ballistic regime. Their derivatives clearly exhibit a number of minima of the dynamic conductance at biases of about 12–13 mV and 3–4 mV, as well as less pronounced features at ~8 mV. The positions of all these features do not constitute a unified subharmonic gap structure. Neither all of them nor any combination of them corresponds to the sequence of Andreev subharmonics at n = 1, 2, …. Therefore, they can describe the properties of different condensates and, according to the formula $V_{n,i} = 2\Delta/en_i$, correspond to the double values of three independent gaps denoted in Fig. 1 as $\Delta_\Gamma$, $\Delta_L$, and $\Delta_S$, or the features $2\Delta_\Gamma$ and $2\Delta_L$ can compose a unified smeared minimum whose width determines the degree of anisotropy of a certain effective gap $2\Delta_L^{eff}$.

We now perform a detailed comparison. The three lower spectra in Fig. 1 are the dynamic conductances (with suppressed exponential behavior) of the arrays LFA11_d2 (two junctions in the stack), LFA11_c (two junctions in the stack), and LFA11_d8 (three junctions in the stack). The stack characteristics were normalized in voltage to the single-junction characteristics [43]. Despite the seeming diversity of the shapes of the features, their qualitative coincidence is achieved after the scaling of biases of these dI/dV spectra by the corresponding integers N. In contrast to the three spectra shown in the upper part of Fig. 1, features near ~8 mV are pronounced and well reproduced, whereas minima at the biases of 10–13 mV are smeared and have small amplitude. Why is such variability manifested? This cannot be explained by the existence of surface states because it is well known that the surface of the LiFeAs compound, which is superconducting in the stoichiometric composition, is chemically pure and is not charged. If the indicated features refer to superconducting gaps opening in different bands, this behavior can be attributed to a significant difference between Fermi velocities in these bands. Because of this difference, the carrier transport conditions for bands with the gaps $\Delta_\Gamma$ and $\Delta_L$ can vary for different microjunctions depending on the properties of the barrier. On the basis of the above consideration, it can be assumed that these gaps are opened in bands most strongly different in the momentum space, i.e., located near the $\Gamma$ and M points of the *k* space.

It can be easily seen that the positions of features marked in Fig. 1 as $2\Delta_S$ coincide and the shapes of these minima in the three upper spectra and the thin solid line for the LFA11_d8 junction from the lower triple are similar. We consider in more detail the shape of the observed features.

Figure 2 shows increased fragments of the experimental spectra from Fig. 1 (points) with a suppressed exponential behavior that contain Andreev features caused by the large gaps $\Delta_\Gamma$ and $\Delta_L$ (LFA11_c and LFA12_f2 junctions, respectively) and from the small gap $\Delta_S$ (LFA11_d8 junction). The doublet character of the feature from the small gap $\Delta_S$ is clearly seen. It involves two minima at biases of ~3.2 and ~2 mV connected by an "arch." Features at biases of 2–2.5 mV are present in all spectra shown in Fig. 1 except for the line for the LFA12_d8 junction, although they are poorly seen in view of the exponential behavior sharply increasing at V → 0. The fitting of the experimental spectrum of the LFA11_d8 junction in Fig. 2 by the theoretical line obtained within the Kümmel–Gunsenheimer–Nikolsky model [39] with the inclusion of anisotropy in the momentum space shows that such a shape of the Andreev reflection is characteristic of the anisotropic gap, which depends on the direction θ in the *k* space as $\Delta_S(\theta) = [1 + A \cos(4\theta)]$, i.e., having four "waves" [45]. The corresponding angular distribution is schematically shown in the inset in Fig. 2 (the position θ = 0 is chosen arbitrarily). Two minima composing the doublet in the conductance spectrum determine the minimum and maximum widths of the small gap by the formula $2\Delta = eV_1$. It can be seen that anisotropy in the *k* space for the small gap is significant, A ≈ 20%, and $\Delta_S^{eff} \approx 1.3$ meV.

Andreev features from the large gap $\Delta_\Gamma$ also have a doublet character as for the small gap. In this case, the doublet consists of two minima at biases of 11.3 and 12.7 mV (see the lower spectrum for the LFA12_f2 junction in Fig. 2). The specificity of the fine structure of $\Delta_\Gamma$ for this and other junctions in Fig. 1 is that their anisotropy is significantly smaller than that for the features of $\Delta_L$ and $\Delta_S$, reaching 8%. Fitting for the LFA12_f2 junction within the Kümmel–Gunsenheimer–Nikolsky model with allowance for anisotropy (see Fig. 2) corresponds to the distribution of the order parameter in the k space $\Delta_\Gamma(\theta) = \{1 + A \cos[4\theta - B \sin(4\theta)]\}$, where the correction $B \sin(4\theta)$ describes the difference between the intensities of the minima of the doublet. The angular distribution of the large gap $\Delta_\Gamma(\theta)$ is shown in the inset in Fig. 2 and corresponds to the values $\Delta_\Gamma^{eff} \approx 11.8$ meV, A ≈ 8%, and B ≈ 35%. Since SNS Andreev spectroscopy is insensitive to the definition of the phase of the order parameter, the angular position θ = 0 was conditionally taken for $\Delta_\Gamma$ and $\Delta_S$ and can be different for these order parameters. We note that the fitting line for $\Delta_\Gamma$ at biases of 9–10 meV does not coincide with the dynamic conductance spectrum. This can be explained by the presence of the Andreev minimum caused by the condensate determined by the order parameter $\Delta_L$. The splitting of the Andreev minimum for the gap $\Delta_L^{eff} \approx 9.4$ meV also indicates its significant anisotropy (about 12%).

At first glance, the doublet minimum in the spectrum of the LFA11_c junction (middle spectrum in Fig. 2) also corresponds to the extended s-type symmetry of the order parameter. However, fitting with allowance for the asymmetry $(1 + A \cos[4\theta - B \sin(4\theta)])$ of the gap even with the correction $B \sin(4\theta)$ does not describe the observed experimental shape of the feature. The strongly asymmetric arch connecting both minima passes too low and does not reach the common horizontal level. Fitting of the experimental spectrum with the different forms of the possible anisotropy of the large gap $(1 + A |\cos(4\theta)|, 1 + 100\% \cos(4\theta),$ and $A \cos(4\theta))$ is also unsatisfactory. Therefore, these minima at the biases of ~7.8 and 10.3 mV cannot be attributed to the splitting of one order parameter; they rather describe the properties of different bands with different gaps. Indeed, the theoretical spectrum (see Fig. 2) corresponding to the case of two independent large gaps opened on different sheets of the Fermi surface reproduces the experimental spectrum with a good accuracy. The weakly asymmetric shape of the minimum marked as $\Delta_L$ (the feature is flatter towards small biases) also implicitly indicates the anisotropy of this gap. Additional Andreev features corresponding to the anisotropy of the gaps $\Delta_\Gamma$ and $\Delta_L$ are strongly smeared possibly because of the weak structural disorder in the region of the LFA11_c junction.

Thus, the dynamic conductance spectra shown in Fig. 1 determine three independent superconducting gaps $\Delta_\Gamma \approx 5.8 \pm 0.7$ meV, $\Delta_L \approx 4.3 \pm 0.5$ meV, and $\Delta_S \approx 1.4 \pm 0.5$ meV. The spread of the values of the gaps $\Delta_L$ and $\Delta_S$ corresponds to their anisotropy. Variation of $\Delta_\Gamma$ at anisotropy that does not exceed 8% is most likely due to imperfection. The characteristic ratios of the BCS theory corresponding to these values are $2\Delta_\Gamma/k_B T_c \approx 7$–9, $2\Delta_L/k_B T_c \approx 5.3$–6.7, and $2\Delta_S/k_B T_c \approx 2$, where $T_c$ is the critical temperature in the bulk, $T_c^{bulk} \approx 16.5$ K.

In order to determine the temperature dependences of the gaps $\Delta_\Gamma$ and $\Delta_S$, the dI(V)/dV characteristic of the single LFA12_b junction was measured in the temperature range $4.2\ K \leq T \leq 16\ K$ (Fig. 3). It is clearly seen that the features from both gaps approach zero and become less intense with an increase in the temperature. The derivative of the current-voltage characteristic is linearized at $T \approx 16$ K. The peculiarities caused by multiple Andreev reflections disappear. This implies the transition of the junction region with the diameter $a \approx 2 \pm 1$ nm to the normal state. The corresponding local critical temperature can differ from the bulk temperature of the sample determined from the temperature dependences of the resistance or magnetic susceptibility. The minimum of the conductance at $V \approx 1.3$ mV is linearly shifted toward smaller biases at T and corresponds to the beginning of the foot. Figure 4 shows the temperature dependences of the gaps $\Delta_\Gamma(T)$ and $\Delta_S(T)$ plotted by the data presented in Fig. 3. It should first be emphasized that the observed features have different temperature dependences. The normalized temperature dependence $\Delta_S(T)\Delta_\Gamma(0)/\Delta_S(0)$ differs from the temperature behavior of the $\Gamma$ gap. This confirms that these features are due to the properties of various superconducting condensates. Furthermore, both temperature dependences are obviously different from the single-gap BCS-like function (see Fig. 4). The dependence $\Delta_S(T)$ begins to decrease already at temperatures of about 4 K and, then, smoothly approaches . The temperature dependence of the large gap generally corresponds to the standard BCS-like curve. However, $\Delta_\Gamma(T)$ at T $\geq$ 4 K is below the last curve and approaches zero almost vertically. Both gaps are closed at the common critical temperature $T_C^{local} \approx$ 16 K.

Such a behavior of the gaps is typical of the proximity effect in the *k* space between the superconducting condensates. It is described by the Moskalenko–Suhl system of gap equations [33], i.e., the two-gap BCS-model. The temperature dependences of the large and small gaps (see Fig. 4) calculated with this system of equations (at the temperature renormalization of the BCS integral necessary for the description of superconductors with the characteristic ratio $2\Delta/k_BT_c > 3.52$) are in good agreement with the experimental dependences. In this case, the ratio $\alpha = N_S/N_\Gamma$ of densities of states in the bands is a free parameter because the dependences obtained from fitting qualitatively reproduce $\Delta_{S,\Gamma}(T)$ in a wide range of $\alpha$ values. To determine the electron-boson coupling constants $\lambda_{ij}$, we took the minimum possible parameter $\alpha_{min}$. Nevertheless, according to our data for LiFeAs, $\alpha_{min} >$ 15. We previously observed such large minimum possible $\alpha$ values for the temperature dependences of the gaps in $MgB_2$, where the $\alpha_{min}$ lays in the interval from 8 to 15.

The relative coupling constants $\lambda_{ij} = V_{ij}N_j$ (where $V_{ij}$ are the matrix elements of the interaction between the $i^{th}$ and $j^{th}$ bands and $N_j$ is the density of states in the $j^{th}$ band), both intraband (i = j) and interband (i ≠ j), were estimated from fitting (see Fig. 4) as $\lambda_{\Gamma\Gamma} : \lambda_{SS} : \lambda_{S\Gamma} : \lambda_{\Gamma S}$ = 0.65 : 0.5 : 0.009 : 0.2 (all constants are normalized to the $\lambda_{SS}$ constant taken as 0.5). The ratio of the densities of states in the bands with $\lambda_\Gamma$ and $\lambda_S$ is estimated as $\alpha = N_S/N_\Gamma \approx$ 22. Such a high ratio is strange because the classical relation implies that $N_1/N_2 = \Delta_2/\Delta_1 \approx$ 4. At the same time, it was assumed in [46] that the relation $\alpha = N_1/N_2 \approx (\Delta_2/\Delta_1)^2$ should be satisfied in the case of the $s^\pm$ symmetry. Then, substituting our experimental values for $\Delta_\Gamma$ and $\Delta_S$, we obtain $\alpha \approx$ 15, which is much closer to a value of 22 determined from the fitting. We note that, in this case, $\lambda_{\Gamma\Gamma}, \lambda_{SS} >> \lambda_{S\Gamma}, \lambda_{\Gamma S}$ and $\beta = \sqrt{\lambda_{\Gamma\Gamma}\lambda_{SS}/\lambda_{S\Gamma}\lambda_{\Gamma S}} \approx 13$. This certainly indicates the dominant role of the intraband interaction, which is inconsistent with the prevalence of interband pairing necessary for the implementation of the $s^\pm$ model [46]. From fitting, we estimated the "solo" characteristic BCS-ratios for the large and small gaps as $2\Delta_\Gamma/k_BT_c^\Gamma \approx$ 7.6 and $2\Delta_S/k_BT_c^S \approx$ 4.6, respectively, where $T_C^\Gamma$ and $T_C^S$ are hypothetical critical temperatures of the respective condensates in the absence of interaction between them. It is obvious that $T_C^S < T_C^{local} < T_C^\Gamma$.

The spectrum of the LFA11_c stack junction measured at various temperatures up to the local temperature $T_c^{local} \approx 16.5$ K is shown in Fig. 5. It clearly exhibits features from three gaps, $\Delta_\Gamma$, $\Delta_L$, and $\Delta_S$, whose temperature dependences are shown in Fig. 6. As was shown above, closely spaced features at V ≈ ± 9 mV cannot be attributed to the split of a single order parameter and are due to two gaps, $2\Delta_\Gamma$ and $2\Delta_L$. The temperature dependence of the difference ($\Delta_\Gamma - \Delta_L$) is also shown in Fig. 6. If these gaps were completely independent or were determined by two regions of the junction with different Δ values in the real space (e.g., owing to the existence of impurities), the dependences $\Delta_L(T)$ and $\Delta_\Gamma(T)$ would approach different $T_c$ values and, thereby, would not have a monotonically decreasing difference $(\Delta_\Gamma - \Delta_L)(T)$. It is noteworthy that, when $\Delta_L(T)$ is normalized to $\Delta_\Gamma(T)$, their temperature dependences are almost the same. According to the Moskalenko–Suhl theory [33], this means that the density of states in the $\Delta_\Gamma$ and $\Delta_L$ bands are approximately the same (α ≈ 1). The complete analogy of deflections on the dependences $\Delta_L(T)$ and $\Delta_\Gamma(T)$ in the three-band approximation indicates that $\lambda_{L\Gamma} \approx \lambda_{\Gamma L} \ll \lambda_{S\Gamma}, \lambda_{SL}$. Consequently, $\Delta_\Gamma$ and $\Delta_L$ can be considered as one effective band with the gap $\Delta_L^{eff} = (\Delta_\Gamma + \Delta_L)/2$. Its temperature dependence is shown in Fig. 7. In addition, the dependences $\Delta_L^{eff}(T)$ and $\Delta_S(T)$ can be fitted by the two-gap model (see Fig. 7). The temperature dependences of the effective large gap and small gap are similar (in contrast to the data shown in Fig. 4). The theoretical dependences $\Delta_i(T)$ reproducing the experimental data with a good accuracy were calculated with the parameters $\alpha_{min} = N_S/N_L^{eff} \approx 16$ and $\lambda_{LL} : \lambda_{SS} : \lambda_{SL} : \lambda_{LS} = 0.6 : 0.5 : 0.011 : 0.17$. In this case, $\Delta_L^{eff}(0) \approx 4.6$ meV, $\Delta_S(0) \approx 1.5$ meV, and $(\Delta_L/\Delta_S)^2 \approx 9.4$ in qualitative agreement with the $\alpha_{min}$ value; nevertheless, the intraband constants λ dominate (β ≈ 13). For the temperature dependence under consideration, the characteristic BCS ratios for the large and small gaps are $2\Delta_L^{eff}/k_B T_c^{Leff} \approx 4.8$ and $2\Delta_S/k_B T_c^S \approx 3.7$, respectively.

Thus, α and relative $\lambda_{ij}$ values for the LFA11_c junction are close to the respective values for the LFA12_b junction. This indicates the good reproducibility of our data obtained by Andreev and intrinsic Andreev spectroscopy. Moreover, this confirms our assumption that the $\Delta_\Gamma$ and $\Delta_L$ condensates hardly interact with each other. The characteristic BCS ratio for the small gap (in the hypothetic case of the complete absence of interband interactions) is $2\Delta_S/k_B T_c^S = 3.7$-$4.6$. As a result, a description is possible within the Eliashberg strong coupling theory [3].

Comparing three gaps determined in this work to the ARPES data (our results are in good agreement with the results reported in [23] and obtained for the same single crystals [34]), we can assume that the large gap $\Delta_\Gamma$ is opened below $T_c$ on the inner hole cylinder near the Γ point of the Brillouin zone, the small gap $\Delta_S$ is opened on the outer cylinder near the $\Gamma$ point, and the middle gap $\Delta_L$ is opened on electron ellipsoids of the M point. This is also in agreement with the theoretical calculations [13, 18], which show that the outer hole cylinder of the Fermi surface is characterized by the highest density of states $N_S$. Taking into account the relations $\lambda_{L\Gamma} \approx \lambda_{\Gamma L} \ll \lambda_{S\Gamma}$, $\lambda_{SL} \ll \lambda_{ii}$ (where i = $\Gamma$, L, S), we can conclude that the interband interaction between hole and electron bands is weak and the intraband interaction is determining.

In summary, the three superconducting gaps, $\Delta_\Gamma$ = 5.1–6.5 meV, $\Delta_L$ = 3.8–4.8 meV, and $\Delta_S$ = 0.9–1.9 meV (at T $\ll$ $T_c$), as well as their temperature dependences, have been directly detected for the first time in a tunneling experiment with the LiFeAs samples with the maximum critical temperature $T_c$. The anisotropy degrees of the order parameters in the k-space have been estimated as <8, ~12, and ~20%, respectively. The characteristic ratios of the BCS theory for these three gaps are $2\Delta_\Gamma/k_B T_c \approx$ 7–9, $2\Delta_L/k_B T_c \approx$ 5.3–6.7, and $2\Delta_S/k_B T_c \approx$ 2, where $T_c$ is the critical temperature in the bulk, $T_c^{bulk} \approx$ 16.5 K.

The signs of the constants $\lambda_{ij}$ cannot be determined from the fit of the temperature dependences of the gaps; consequently, it is impossible to verify the $s^{++}$ or $s^{\pm}$ symmetry. At the same time, comparing $2\Delta_i$ to the experimentally determined energy of magnetic resonance for LiFeAs from [8], we can find that $E_{res} \approx 0.8\Delta_S \approx 2\Delta_L \approx 3\Delta_\Gamma$ at $E_{res}/k_B T_c \approx 5.5$. Therefore, it can be implicitly concluded (see [7]) that the wavefunctions of the order parameters $\Delta_L$ and $\Delta_S$ are in one phase ($s^{++}$). Only the order parameter $\Delta_\Gamma$, which is formed on the inner hole cylinder near the $\Gamma$ point of the Brillouin zone, can be in antiphase.

We are grateful to Prof. Ya.G. Ponomarev for support, to S.V. Borisenko, P.I. Arseev, N.K. Fedorov, and I.A. Devyatov for stimulating discussions, and particularly to S. Wurmehl, and B. Büchner for samples provided for investigations. This work was supported by the Council of the President of the Russian Federation for Support of Young Scientists and Leading Scientific Schools (project no. MK-3264.2012.2), by the Russian Foundation for Basic Research (project no. 13-02-01451-a), by the ERA.Net RUS program (grant nos. STProjects-245 FeSuCo and 12-03-91674-ERA), and by Deutsche Forschungsgemeinschaft (grant nos. BE1749/13 and BU887/15-1).

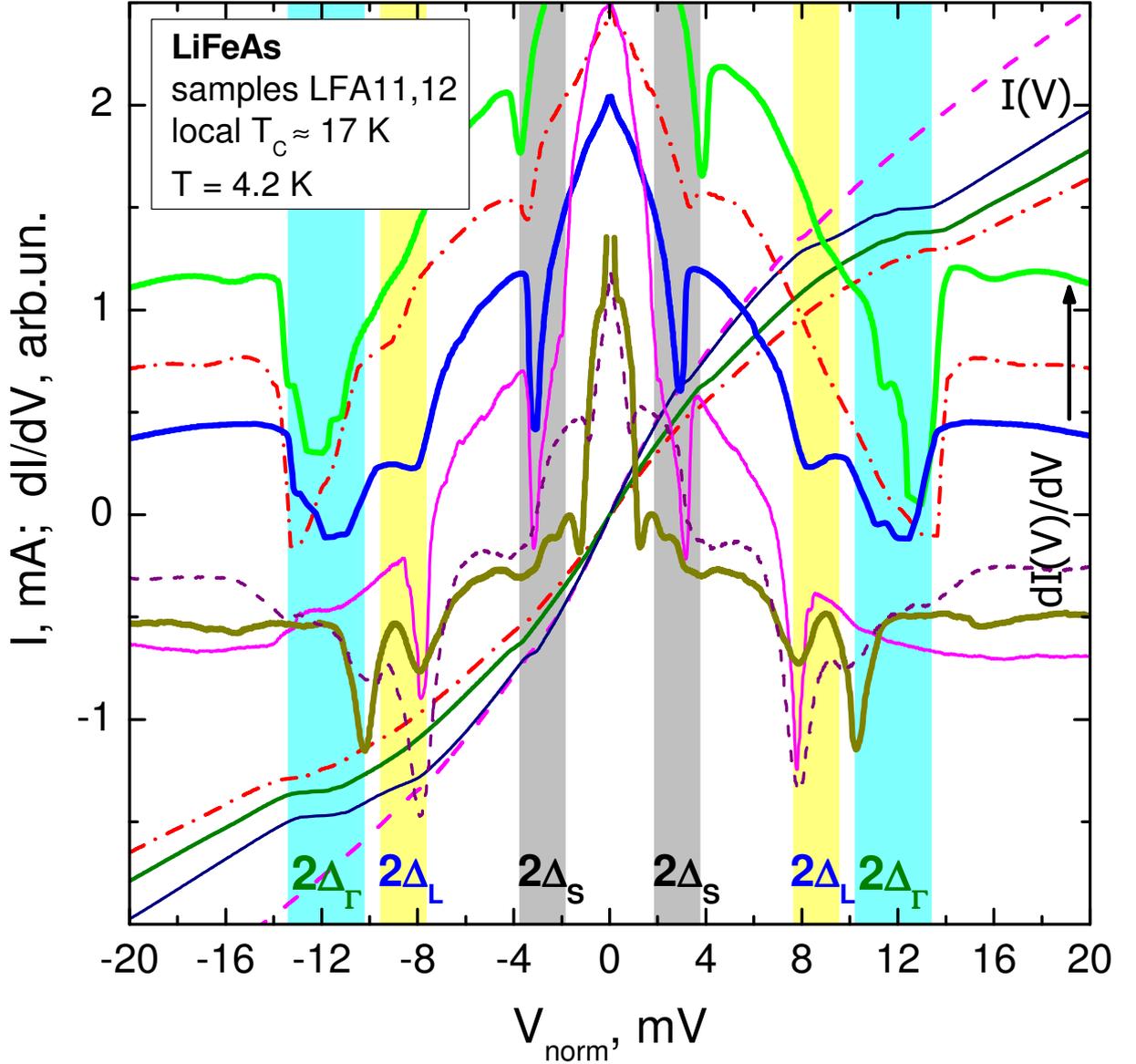

**Fig. 1.** Current-voltage characteristics I(V) and dynamic conductance spectra dI(V)/dV for the single SNS Andreev junctions (thick black lines) LFA12_d8, (dash-dotted lines) LFA12_b, and (upper gray line) LFA12_f2 obtained by the successive mechanical readjustment, as well as stack junctions (dashed lines) LFA11_d2 (two junctions in the stack), (gray lower line) LFA11_c (two junctions in the stack), and (thin solid line) LFA11_d8 (three junctions in the stack). The stack characteristics were normalized to a single junction. The data were obtained at T = 4.2 K. The local critical temperatures of the junctions are approximately the same, $T_c \approx$ 16–17 K. The vertical regions mark the positions of Andreev peculiarities corresponding to the superconducting gaps $\Delta_\Gamma \approx 5.8 \pm 0.7$ meV, $\Delta_L \approx 4.3 \pm 0.5$ meV, and $\Delta_S \approx 1.4 \pm 0.5$ meV. The spread of $\Delta_{L,S}$ values corresponds to their anisotropy.

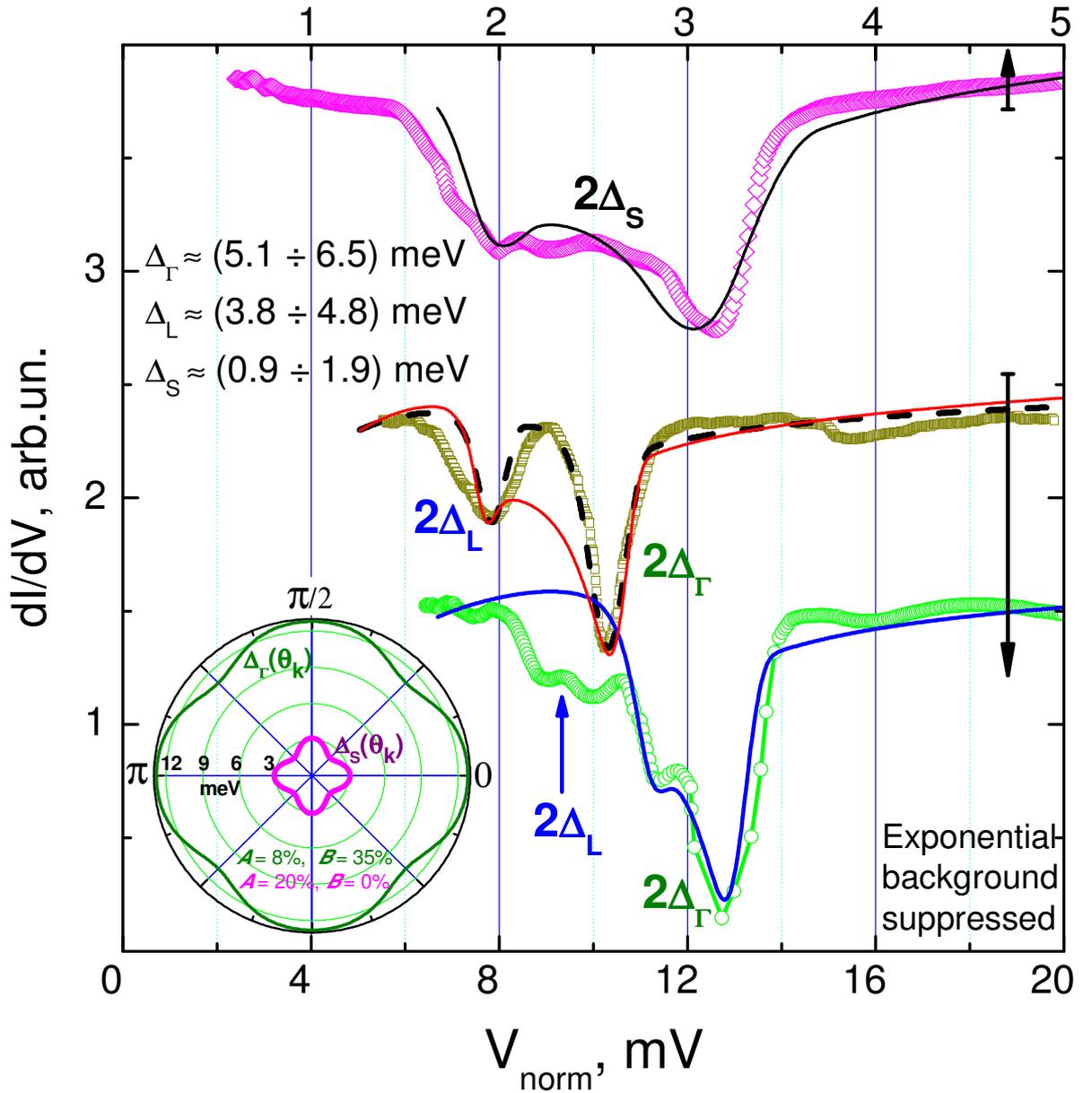

**Fig. 2.** Increased fragments of the experimental spectra from Fig. 1 (points) with a suppressed exponential behavior that contain Andreev features from the large gaps $\Delta_\Gamma$ and $\Delta_L$ (LFA11_c and LFA12_f2 junctions, lower dark and gray lines, respectively, refer to the lower bias axis) and from the small gap $\Delta_S$ (LFA11_d8 junction, upper line refers to the upper bias axis). The experimental spectra are fitted by the theoretical lines obtained within the Kümmel Gunsenheimer–Nikolsky model [39] (solid lines) for the cases of the anisotropy of the gap in the k-space and (dashed line) for the case of two independent gaps (LFA11_c junction). The inset shows the distributions of the gaps $\Delta_\Gamma$ and $\Delta_S$ in the direction of the k-space. The angular position $\theta = 0$ is taken for the gaps $\Delta_\Gamma$ and $\Delta_S$ conditionally and can be different for these order parameters.

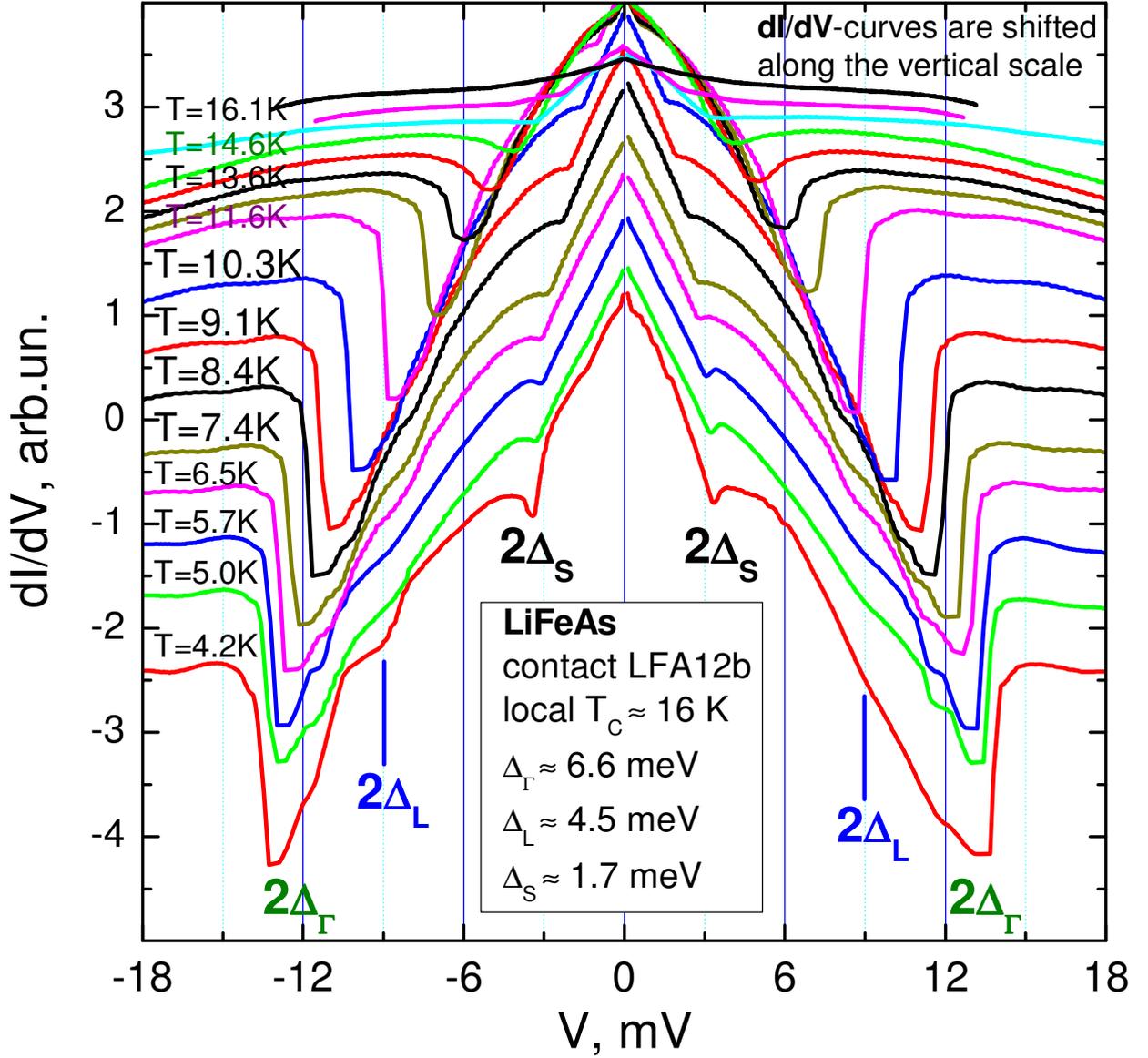

**Fig. 3.** Dynamic conductance spectra of the LFA12_b junction measured in the temperature interval 4.2 K ≤ T ≤ $T_c^{local}$ ≈ 16 K. The characteristics are vertically shifted for convenience. The positions of Andreev reflections are marked as $2\Delta_\Gamma$, $2\Delta_L$, and $2\Delta_S$.

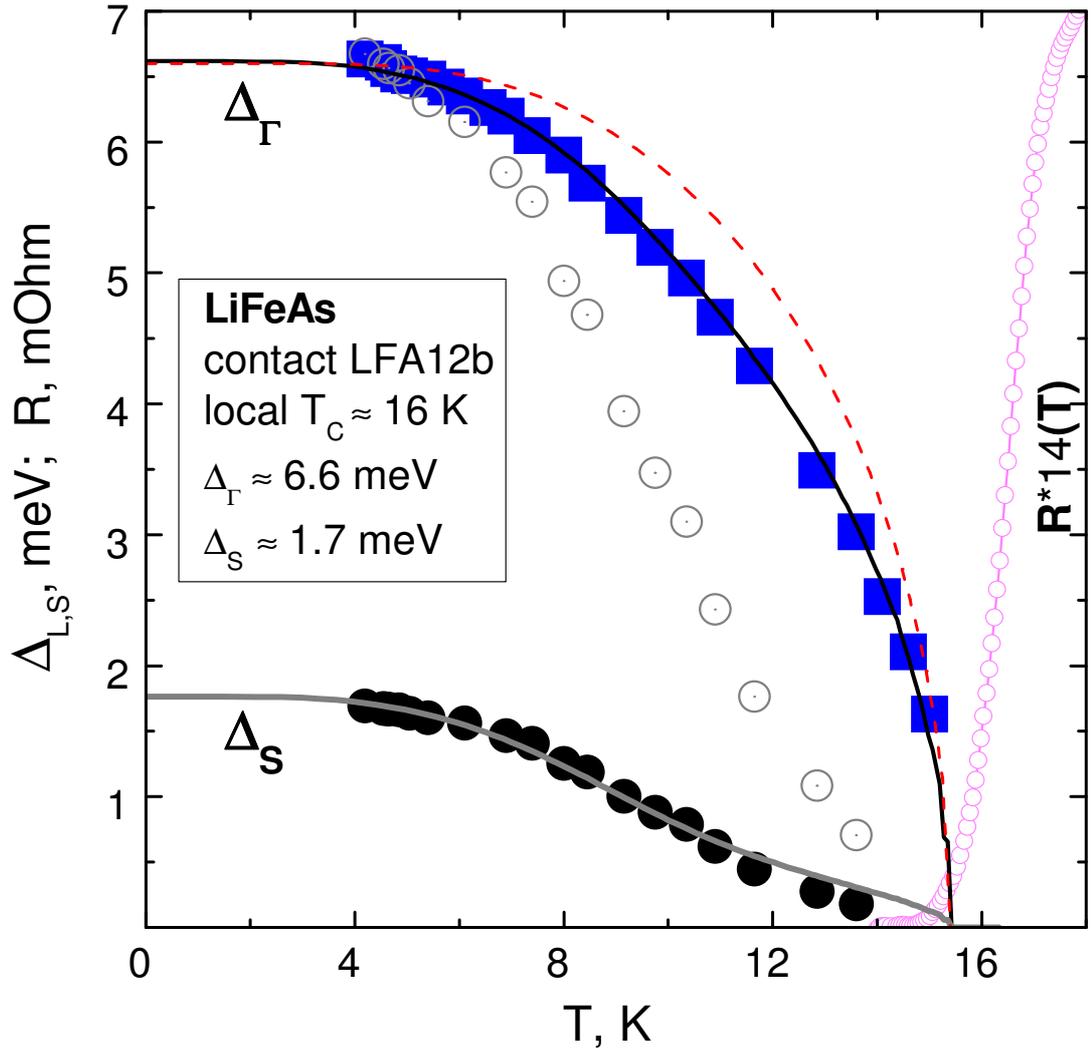

**Fig. 4.** Temperature dependences of the gaps (squares) $\Delta_\Gamma(T)$ and (closed circles) $\Delta_S(T)$ plotted according to the data presented in Fig. 3. For comparison, open circles (dark and light, respectively) represent the normalized dependence $\Delta_S(T)\cdot\Delta_\Gamma(0)/\Delta_S(0)$ and resistive transition of the LFA12 sample. The solid lines are the theoretical temperature dependences of the gaps calculated within the two-gap BCS Moskalenko–Suhl model [33]. The dashed line corresponds to the standard single-gap BCS-like function.

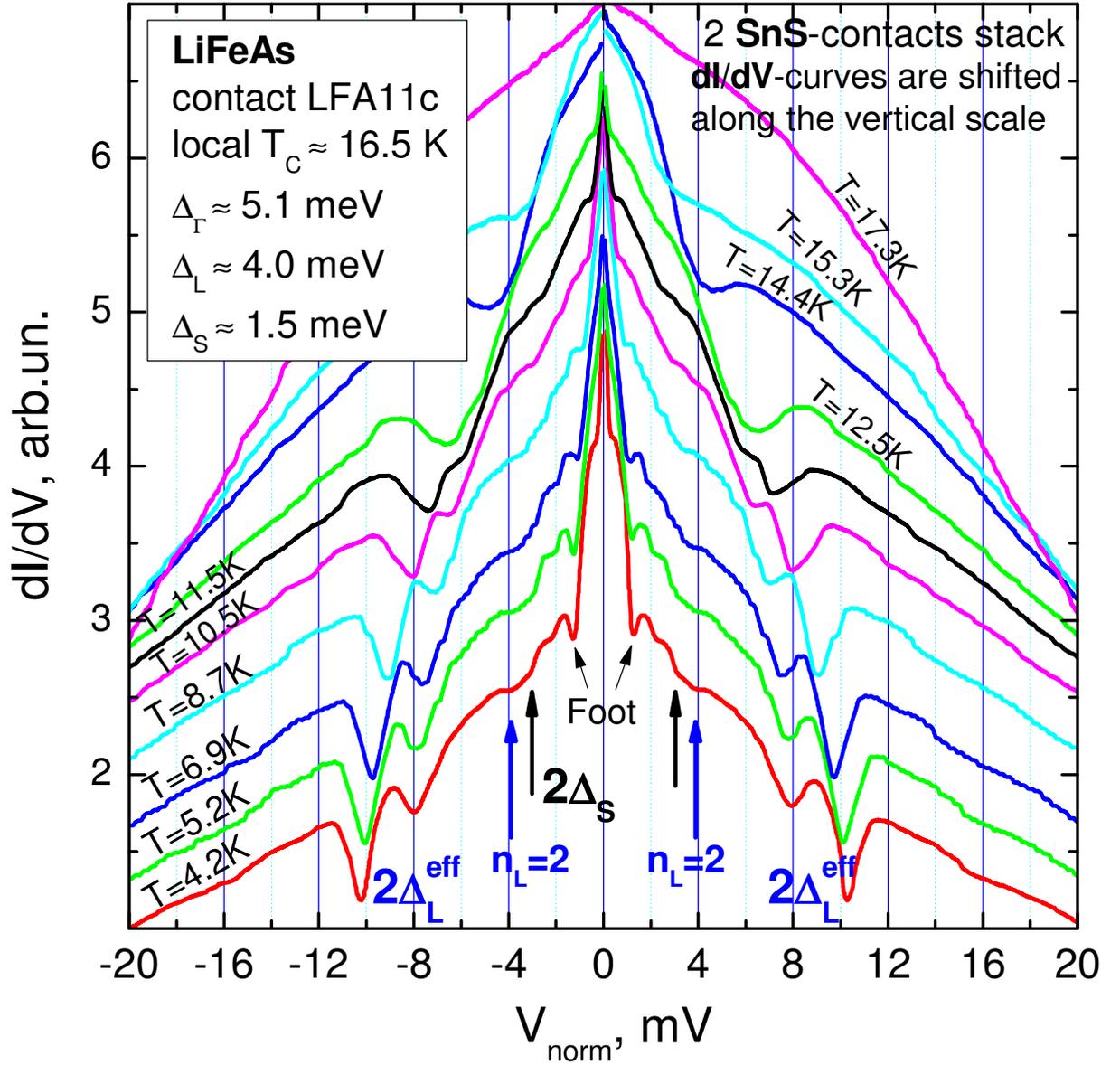

**Fig. 5.** Normalized dynamic conductance spectra of the LFA11_c junction obtained in the temperature interval 4.2 K ≤ T ≤ $T_c^{local}$ ≈ 16.5 K. The characteristics are vertically shifted for convenience. The positions of Andreev reflections for large gaps $\Delta_\Gamma$ and $\Delta_L$ are marked as $2\Delta_L^{eff}$ and $n_L = 2$, respectively, and for the small gap, as $2\Delta_S$.

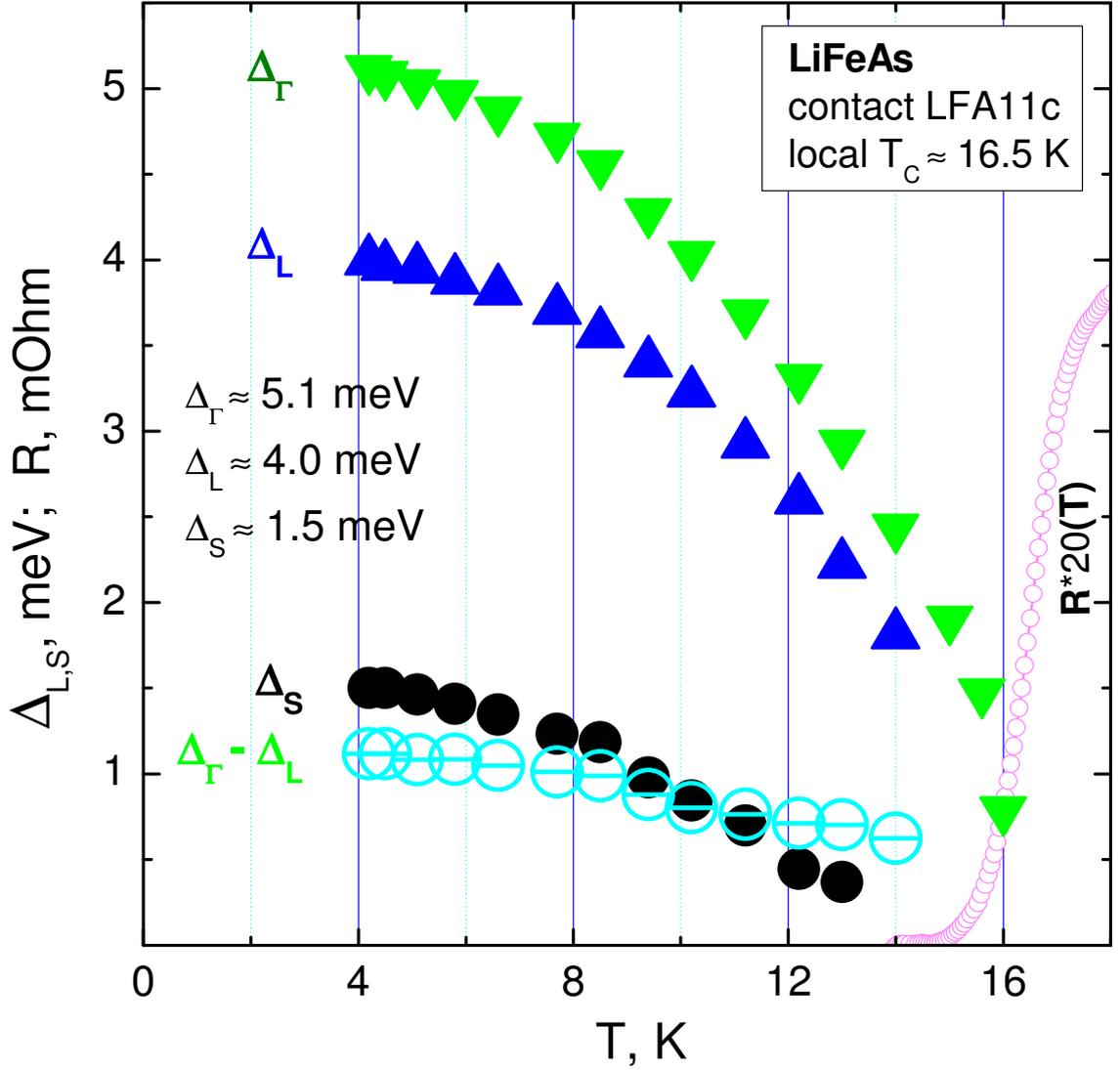

**Fig. 6.** Temperature dependences of the gaps (inverted triangles) $\Delta_\Gamma(T)$, (direct triangles) $\Delta_L(T)$, (open crossed circles) their difference $\Delta_\Gamma - \Delta_L$, and (closed circles) $\Delta_S(T)$ plotted according to the data presented in Fig. 5. For comparison, open circles represent the resistive transition in the LFA11 sample. The behavior of the difference $\Delta_\Gamma - \Delta_L$ differs from that of $\Delta_S(T)$. The dependences of the order parameters $\Delta_\Gamma(T)$ and $\Delta_L(T)$ are identical up to a linear coefficient and approach the same $T_c$ value.

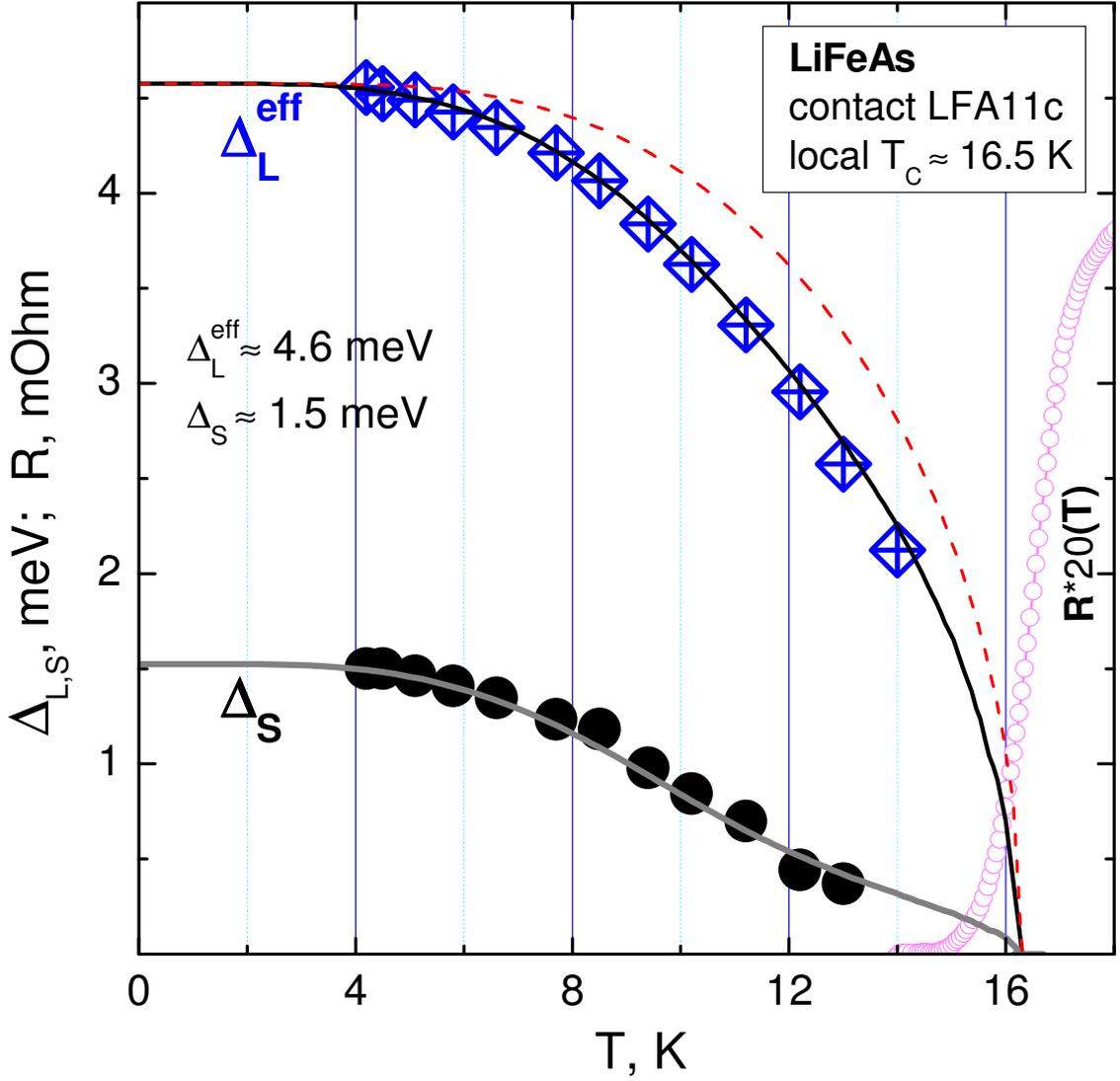

**Fig. 7.** Temperature dependences of the effective large gap (crossed diamonds) $\Delta_L^{eff} = (\Delta_\Gamma + \Delta_L)/2$ and (closed circles) $\Delta_S(T)$ plotted according to the data presented in Fig. 5. For comparison, open circles represent the resistive transition in the LFA11 sample. The solid lines are the theoretical temperature dependences of the gaps calculated within the two-gap BCS Moskalenko–Suhl model [33]. The dashed line corresponds to the standard single-gap BCS-like function.